# Optical detection of the spatial structural alteration in the human brain tissues/cells and DNA/chromatin due to Parkinson's disease

Fatemah Alharthi[1], Dhruvil Solanki[1], Ishmael Apachigawo[1], Jianfeng Xiao[2], Mohammad Moshahid Khan[2,*], Prabhakar Pradhan[1,*]

[1]Department of Physics and Astronomy, Mississippi State University, Mississippi State, MS, USA, 39762
[2]Department of Neurology, University of Tennessee Health Science Center, Memphis, TN, USA, 38103

*Corresponding Authors:* MMK and PP

**Abstract:** Parkinson's disease (PD) is considered one of the most frequent neurological diseases in the world. There is a need to study the early and efficient biomarkers of Parkinson's, such as changes in structural disorders like DNA/chromatin, especially at the subcellular level in the human brain. We used two techniques, Partial wave spectroscopy (PWS) and Inverse Participation Ratio (IPR), to detect the changes in structural disorder in the human brain tissue samples. It was observed from the PWS experiment that there was an increase in structural disorder in Parkinson's disease tissues/cells when compared to normal tissues/cells using mesoscopic light transport theory. Furthermore, the IPR experiment also showed DNA/chromatin structural alterations that have the same trend and support the PWS results. The increase in mass density in the nuclei components, such as DNA/chromatin, can be linked to the aggregation of alpha-synuclein in the substantia nigra of the brain. This protein deposition is considered a significant cause of neuronal death in the brains of PD patients. We also did a histological analysis of brain tissues, which supports our results from dual photonics techniques. The results show that this dual technique is a powerful approach to detect the changes. Our results highlight the potential of the parameter, related to the structural disorder strength, as an efficient biomarker for PD progress, paving the way for research into early disease detection.

## 1. Introduction

Parkinson's disease (PD) is a complex degenerative age-related neurological disease that leads to a movement disorder[1,2]. A slow decline in motor and non-motor functions characterizes the syndrome. The condition results from nerve cells that are damaged or die in a region of the midbrain called the substantia nigra pars compacta (SNc), which is an important brain area that produces dopamine [3]. The reduced dopamine levels have resulted in various disorders like Parkinson's disease[4,5]. With the advancement of the ailment, a considerable reduction of fifty to ninety dopaminergic cells in the substantia nigra is commonly observed, leading to the manifestation of various signs and symptoms [6].

In addition, protein deposition is considered the first significant cause of neuronal death in the brain of Parkinson's disease patients[7]. Among PD patients, alpha-synuclein is dramatically increased in the brains, and as this protein is insoluble, it congregates to form Lewy bodies in neurons[8]. These structural alterations appear in the brains of Parkinson's patients many decades before the critical symptoms appear [9]. Moreover, these abnormal changes occur at the nanoscale level, i.e., nanoarchitecture within the diseased cells/tissues that cannot be observed with conventional optical methods due to diffraction-limited resolution. Structures below 200 nanometers, such as ribosomes, DNA, proteins, lipids, etc, are not clearly visible under traditional optical techniques due to the diffraction limit.

We have applied a mesoscopic physics analysis to analyze the nanoscale structural changes in PD patients' brains, ranging from nano-to-submicron-scales. This fundamental principle is based on the mesoscopic light transport theory [10–12]. In this type of analysis, the reflection signals coming from the samples result from interferences of light waves, which occur due to refractive index changes inside the sample. These reflected signals are hypersensitive to any length scale of refractive index variations.[13,14]. As a result, the spectra of the 1D scattering signals provide detailed information for any particle, as well as particles with scale lengths even below the wavelength (200 nm). We recently introduced mesoscopic physics-based elastic scattering technique, Partial Wave Spectroscopy (PWS) [13,15]. Using this innovative optical approach, we can acquire 1D reflected waves from different sections of the sample or the medium[16]. Moreover, the technique allows for quantifying changes in the nanoarchitecture in cells/tissues and determining the statistical properties of refractive index variations ($dn(r)$) at the nanoscale [11]. In our previous publications, we reported that PWS was able to detect subtle structural alterations that occur in brain tissues due to chronic alcoholism and the abnormalities changes that happen in prostate carcinogenesis[17,18].

Conversely, the changes in the brain cell due to Parkinson's disease can vary in scale from bulk structural changes to alterations in the molecular-specific structures in cells at the nanoscale. Using the sensitive imaging PWS technique, we can determine that the structures of brain cells/tissues are altered at the nanoscale due to PD[15,19]. In addition, when the disease is in its early stage, nanoarchitecture changes occur in the cell due to the rearrangement of specific molecules such as DNA/chromatin and other molecules, which is the second focus of this study. Accordingly, it is assumed that a system's physical state can be explained by probing its molecular structure alterations within the cells. The strategy of investigating molecular-specific structural alterations using the IPR technique has been developed in recent years to examine nanoscale alterations in molecules within the cells due to various diseases and abnormalities [20–22]. In the IPR, an optical lattice is made using pixel intensities of the confocal image, which is used to solve for the eigenvalues and eigenfunctions. Then, the light localization characteristics of the target molecules specific inside the cells are calculated using the eigenfunctions to calculate the average of the IPR, $\langle IPR \rangle$ and its standard deviation, $\sigma(\langle IPR \rangle)$. As a result of this technique, advanced cancer within the cells can be quantified, particularly when determining the progression of malignancy and the effect of anticancer medications [20,23,24].

In this paper, we report that we detected those changes earlier in human cells/tissues for carcinogenesis using the IPR and PWS, dual optical spectroscopy techniques that are translatable to detect Parkinson's disease. Our results also indicate that optical nano-cytology PWS and IPR methods may promise that early PD identification can lead to therapy advances that delay the neurodegenerative process. In addition, this approach can serve as a minimally intrusive screening test for PD.

The methodology and principle of the dual robust optical/photonics spectroscopic technique, PWS and IPR are explained in detail elsewhere. Thus, here we will present the study's results and conclusions.

## 2. Materials and Methods

### 2.1 Sample Preparation Optical Experiments

Michigan Brain Bank, Ann Arbor, MI, USA and the University of Iowa, Iowa City, IA, USA Deeded body program provided the human brain tissue samples of PD and non-PD patients for the study. The University of Tennessee Health Science Center Institutional Review Board (IRB #20-07595-NHSR; Exempt Application 874552), Memphis, TN, USA approved the study. All the standard ethical procedures were followed throughout the study. Human samples were handled following personal protection safety procedures. The mid-brain blocks, which also include Substantia Nigra (SN) were embedded in paraffin wax and cut a thickness of 6.0 µm on slides for the study of PD pathology.

## 2.2 Sample Preparation for PWS Experiments

The fixed samples were embedded by applying a standard protocol in paraffin wax, a common technique in histology that allows for thin sectioning of the tissue. Then, the tissue samples was sliced into thin sections, typically around *5 µm* thick, using a microtome. These sections were then mounted onto glass slides. Finally, the slides were subjected to PWS measurements, which involved illuminating the tissue with light and measuring the scattering of the light waves. The resulting data was used to analyze the brain tissue structure and identify any abnormal cellular changes.

## 2.3 Sample Preparation for IPR Experiments

In this part, the same human brain samples subjected to PWS measurements were used for IPR analysis. We used a nuclear dye DAPI, which primarily recognizes DNA and chromatin, to investigate the spatial molecular structure of DNA in human brain cells. The tissues were plated on sterile glass slides. The tissues were briefly rinsed for 5 minutes at room temperature inside a phosphate-buffered saline (PBS) containing 2-4% paraformaldehyde. Once 5 minutes of PBS washes were done three times, Prolong Diamond Antifade Mountant, which contains the blue DNA stain DAPI was used to stain glass slides. Nuclei are easily distinguished in fluorescence microscopy when DAPI is present in the sample.

## 2.4 Partial Wave Spectroscopy

### 2.4.1 Optical Setup

A detailed description of the PWS setup has already been discussed in our previous work[11,13,18]. However, for the sake of continuity, a brief description is given in this section. The PWS setup comprises a broadband light source from a Xenon lamp (150 W) (Newport Corporation, Irvine, CA, USA). 4f system of lenses collimates the light and then goes onto the right-angle prism (BRP) and a 50:50 beam splitter, which directs the light towards the objective lens with objective 40x (NA-0.65). The sample gets illuminated by a Kohler illumination, which is kept on a XYZ motorized stage (X-Y axis: 40 nm, Z axis: 100 nm, Zaber Technologies, Vancouver,

BC, Canada) to achieve finer focusing. The backscattered light from the sample gets directed to a liquid crystal tunable filter (Kurios LCTF) (Thorlabs, Newton, NJ, USA) by plano-convex lens (f=10cm) which then is focused to a CCD for image acquisition in a visible spectrum at each wavelength from 450-700 nm. This setup measures the structural disorder, $L_{d\text{-}PWS}$, of the sample at each pixel (x,y) using the backscattered signal at different wavelengths.

### 2.4.2 Calculation of strength of structural disorder using Mesoscopic theory of transport light

Once the reflected signal gets recorded, the fluctuating part is extracted from it and processed to remove the noise with high frequency by passing through a low-pass Butterworth filter. A 3D data cube is formed of the reflected intensity $I(x,y,\lambda)$, where (x,y) is the spatial position of the sample and $\lambda$ is the wavelength at which the spectra were recorded. The theory of mesoscopic light transport is applied to the reflected signal coming from the sample because it is a weakly disordered media[12]. This theory is also applicable to the transport of electrons and light in a dielectric media. The incoming reflected signal is divided into many 1D channels of size 200x200 nm by a quasi-1D approximation[10,25]. The backscattered reflection intensity propagating in the 1D channel is sensitive to fluctuations in refractive index in any length scale. To quantify all these and to evaluate how the structure of the cells/tissues changes, a statistical parameter structural disorder strength $L_{d\text{-}PWS}$ was found. We can calculate it using the *RMS* value of reflected intensity $<R(k)>_{rms}$ and the spectral autocorrelation of the reflection intensity $C(\Delta k)$ at each spatial position[19,26–28].

$$L_d = \frac{B n_0^2 \langle R \rangle}{2k^2} \frac{(\Delta k)^2}{-\ln(\langle C(\Delta k) \rangle)} \quad (1)$$

In equation 1, $B$ is a calibration constant, $n_0 = 1.0$, $k$ is the wavenumber, $\frac{(\Delta k)^2}{-\ln(\langle C(\Delta k) \rangle)}$ can be found by getting a linear fit of $ln(C(\Delta k))$ vs $(\Delta k)^2$.

Structural disorder strength tells us how the structure of the cells/tissues changes as the disease progresses[16]. As the disease progresses, the refractive index fluctuations increase, which can be linked to mass density changes in tissues/cells. $L_d$ can also be written in another form $L_d = <\Delta n^2> \times l_c$, where $<\Delta n^2>$ is the variance of the refractive index fluctuations and $l_c$ is the correlation length.

### 2.5 Confocal Microscopy and Inverse Participation Ratio

Biological samples stained with DAPI were imaged using Zeiss710 confocal microscope (Carl Zeiss Microscopy, Jena, Germany). Images were taken in Z-stack mode above and below the plane of the cell nucleus's central plane. Cells that show the most change in the formation of DNA and chromatin were selected. They were analyzed using software like ImageJ v1.54c (National Institute of Health, Bethesda, MD, USA) and MATLAB.

We use another parameter called the Inverse Participation Ratio (IPR) to analyze the confocal images. This approach is used to statistically study the weakly disordered medium's light localization characteristics by calculating the eigenvalue of the eigenfunction of an optical system. The eigenfunction depends on the intensity of the incoming light waves, which is because of multiple interferences[20,29], and helps us to measure the structural disorder. IPR is evaluated using eigenfunction $IPR = \int |E(r)|^4 dr$, where E is the eigenfunction of the Hamiltonian.

Scattering from cell components such as DNA, lipids, and proteins shows the relationship between the refractive index (n) and the mass density inside the cells. Intensity coming from the voxel is $I(x,y) = I_0 + dI(x,y)$, where $I_0$ is the average intensity and $dI$ is the intensity fluctuations. This can be linked to refractive index, and further to mass density by $n(x,y) = n_0(x,y) + dn(x,y) = \rho_0 + \beta \times d\rho(x,y)$, where $n_0$ is the mean refractive index, dn is the refractive index changes, $\rho_0$ is the mean mass density, and $d\rho$ is the fluctuation in the mass density[30]. Optical potential can be defined using the above properties,

$$\varepsilon_i(x, y) = \frac{dn(x,y)}{n_0} \propto \frac{dI(x,y)}{I_0} \tag{2}$$

where, $\varepsilon_i$ is the optical potential at the pixel position (x,y) of the micrograph acquired from the confocal microscope.

Further on, the tight binding model (TBM) is used to find the Hamiltonian matrix,

$$H = \sum \varepsilon_i |i><i| + t\sum \langle ij \rangle (|i><j| + |i><j|) \tag{3}$$

$\varepsilon_i$ is the optical potential at the $i^{th}$ lattice site, |i> and |j> are the photon's state at the ith and jth lattice site, and t is the inter-lattice. Mean IPR is further calculated as

$$\langle IPR \rangle_{L \times L} = \frac{1}{N} \sum_{i=1}^{N} \int_0^L \int_0^L E_i^4(x,y) dx dy \tag{4}$$

$E_i$ in the above equation is the Hamiltonian's eigenfunction with the optical lattice size of L x L. $N = (La)^2$ [a = dx = dy] is the total eigenfunction. The structural disorder is quantified inside the biological cells with refractive index by $L_{d\text{-}IPR} = <\Delta n> \times l_c$.

$$\text{mean}\langle IPR(L)\rangle_{L\times L} \propto L_d = <\Delta n> \times l_c \qquad (5)$$

$$STD\langle IPR(L)\rangle_{L\times L} \propto L_d = <\Delta n> \times l_c \qquad (6)$$

## 3. Results and Discussion

### 3.1. PWS Analysis of Brain Tissues

To prove the proficiency of the present PWS method in identifying nano-architectural changes in some diseases and modifications of these diseases on the structural changes of the cells/tissues, we performed our experiments on human brain tissues of PD patients. Our study centered on Parkinson's disease (PD), a major public health issue in the United States. In addition, the growing number of people with PD in the United States is burdening the healthcare system. Fig.1(a) and (b) illustrate bright-field images or microscope images of human brain cells/tissues obtained from a control and a PD patient, respectively. These images seem to be indistinguishable microscopically, thus indicating no noticeable changes at microscopic length scales (200 nm). However, when the color maps of the spatial structural distribution of PD, $L_{d-pWS}$ are plotted [Fig.1(a') and (b')], some regions appear with a red color, suggesting higher disorder strength $L_d(x,y)$, resulting from the nanoscale abnormalities of the brain tissues obtained from PD patients compared to the control. This indicates the nanoscale sensitivity of the PWS technique.

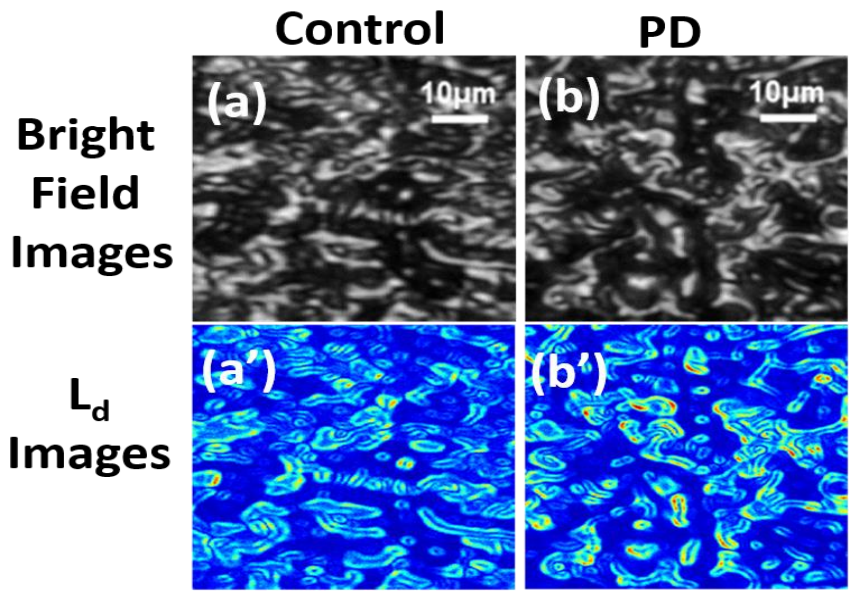

**Figure 1:** Representative bright field microscope and images $L_d$ of human brain cells and tissues from control and PD patients. (a) and (b) are bright field microscopic images of brain cells and tissue from control and PD patients, respectively. (a') and (b') are representative PWS-produced color maps of $L_d$ for the brain cells and tissues of the control and PD patients. Although the microscopic images of the diseased cells and tissues and the healthy ones (control) are indistinguishable, the disorder strength ($L_d$)) significantly increased in PD patients compared to the control.

For statistical comparison, Fig. 2 (a) and (b) show the PWS analysis of human brain tissues. In this figure, the corresponding bars in the graph indicate the average and standard deviation of $L_d$ of human brain tissue for patients with PD and healthy individuals along with their corresponding standard error. $L_{d\text{-}PWS}$ was found to be increasing in PD patients compared to their control, which is shown in bar plots. The percentage changes in the average and standard deviation of $L_d$ relative to the control human brain tissue for PD patients are approximately 26.8% and 27.5 %, respectively. The effect of Parkinson's disease on the human brain results in spatial mass density alterations in cells/tissues due to changes in and rearrangements of the cell-internal components such as DNA and RNA. These fluctuations in mass density result in fluctuations in the refractive index. Parameter dn can be measured and expressed as the degree of disorder strength $L_{d-PWS}$. This effective and powerful biomarker/numerical index can more precisely quantify nanoscale alterations and differentiate between healthy and diseased cells and tissues. Due to the sensitivity of PWS microscopy to detect these nanoscale alterations within cells and tissues, this technology can offer a noninvasive method for early PD risk classification and provide a way to obtain predictive information on the disease.

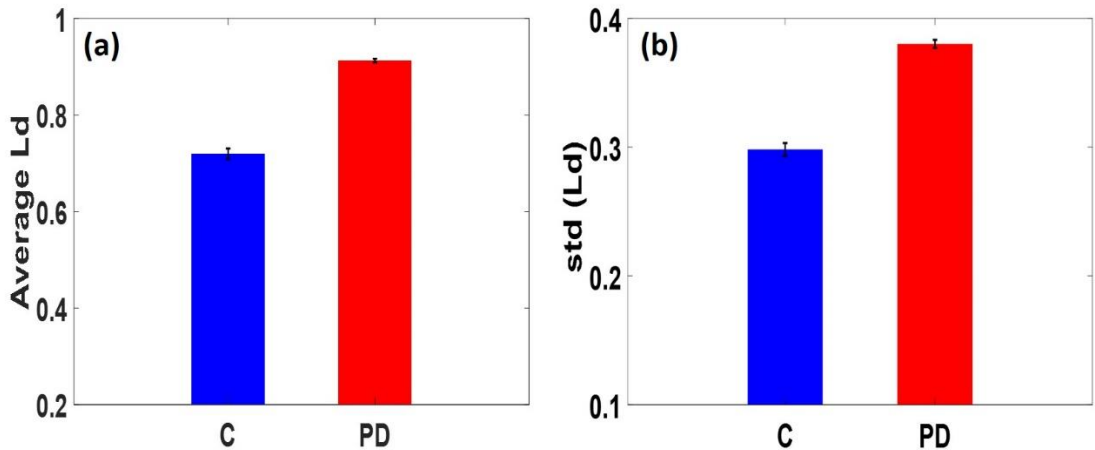

**Figure 2:** The bar plots (a, b) represent the mean and std of $L_d$ calculated from human brain tissues of Control and PD patients, respectively. The $L_{d-PWS}$ results show that the average $L_d$ value correlates with the PD of the tissues. The average $L_d$ value of Parkinson's disease (PD) increases by 26.8%, and the standard deviation of $L_d$ increases by 27.5 % (P- value<0.05), relative to control (C). This confirms that the PD has caused a nanoscale structural alteration of the patient's brain.

### 3.2. IPR Analysis on Brain Tissues

Using the procedure outlined in Section 2.5, images captured from a confocal microscope of the samples were investigated for disorder inside the DNA, and the results were statistically assessed. The degree of disorder strength $L_{d-IPR}$ equivalent to IPR values, were calculated for the length scale L=0.4 µm. Then, 12–15 cells from each of the two categories were examined after calculating the average IPR values for 5-8 confocal images of each single cell nuclei. Fig. 3(a), and (b) show representative confocal images of the 1) normal human brain, DAPI stained and 2) PD of the human brain DAPI stained cell's nucleus, respectively. The images shown in Fig. 3(a') and (b') are the corresponding IPR images that were acquired with a sample length of L=0.4 µm. These IPR images illustrate the distribution of the IPR values are the degree of structural disorder $L_d$, determined in the two cells' nuclei sample area of 0.4 x 0.4 µm². Fig.3(a) and (b) show confocal images of the nuclei of healthy and diseased cells, respectively. The IPR images reveal more pronounced hot spots (red spots) regions in the colormap in the PD cell nucleus than the healthy cell nucleus, suggesting a higher structural disorder.

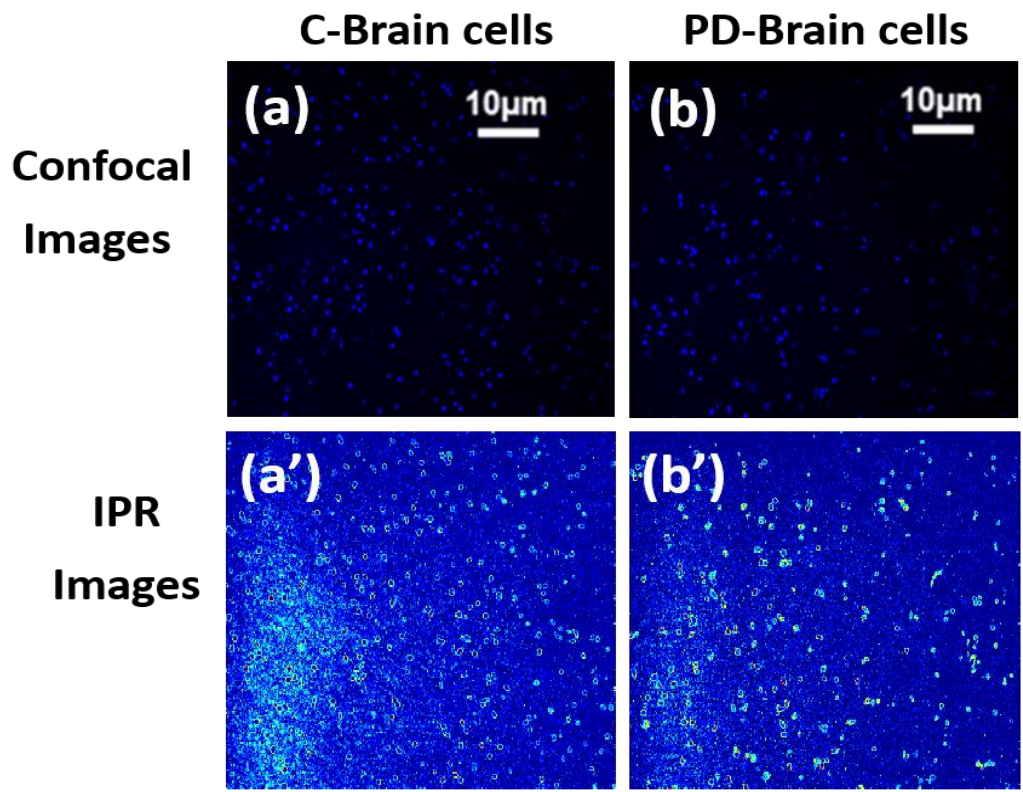

**Figure 3:** Representative confocal images of DAPI stained and IPR images of human brain cells from control and PD patients. (a) and (b) are representative confocal images of normal cell nuclei and PD cell nuclei, DAPI stained, respectively, and (a') and (b') are their following disorder strength (IPR) images with L =0.4 µm.

A comparison of the mean and *std* of the structural disorder, $L_{d-IPR}$ or <IPR> values between the normal human brain cells and the PD cells are shown in Fig. 4 (a) and (b). The results of the mean and STD of IPR analysis demonstrate that PD cells' structural disorder of the human brain is significantly higher than the control cells. The percentage difference between the PD cells relative to the normal cells for the mean and STD of the IPR value is 53% and 2.4%, respectively. This result indicates that the STD of IPR values does not change much compared to the mean of IPR values, which varies significantly. The PD brain cells' higher IPR values imply that their nuclear DNA and chromatin have more nanoscale structural abnormalities/damages than normal cells. The unfolding of alpha-synuclein protein within the nuclei during Parkinson's disease may be responsible for the increased structural disorder in the DNA of the afflicted cell.

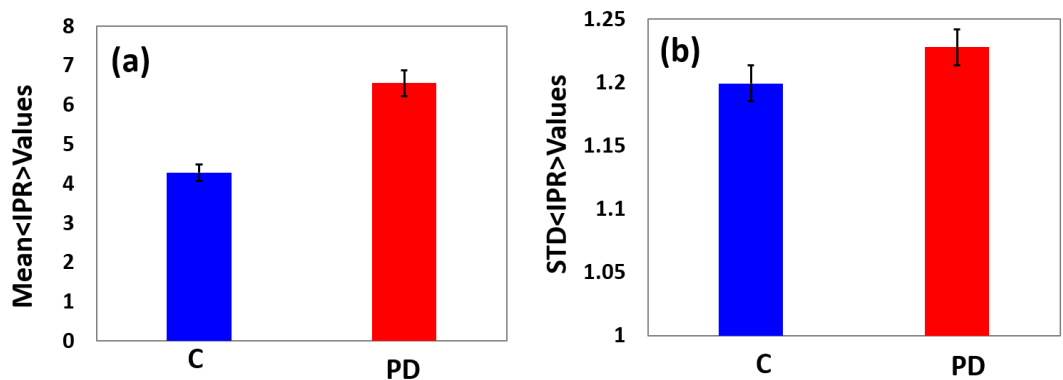

**Figure 4:** The IPR analysis of the human brain cell's nuclei, DAPI stained (a) and (b) are the bar plots for mean and standard deviation of IPR values (n=10-15 cells, 5-11 images per cell, ~6 sets) for the control and PD cells nuclei at sample length L= 0.4 µm; p-value < 0.05 for each pair. The result shows that the mean and STD of the IPR values of PD cell samples are higher than the control. The percentage increase is 53% and 2.4%, respectively.

### 3.3. Histology and Immunofluorescence study of human PD and non-PD brains:

*3.3.1. Histology of human PD and non-PD brains:*

Immunofluorescent staining was carried out which is described here[31]. Coronal sections from the midbrain level were chosen for the immunohistochemical analysis of α-Synuclein. Later, they were pretreated with 10mM citric acid buffer whose pH is 6.0 for antigen removal and then 3 PBS washes. They were then blocked using 5% bovine serum albumin (BSA; Sigma Aldrich #A7906) for 1 hour. These sections were incubated with primary antibody to target phosphor α-Synuclein (Ser129) (BioLegend) for a night at 4°C. After 3 PBS washes, they were incubated at room temperature for 1 hour using Universal biotinylated secondary antibody (Vector laboratories). Again, sections were washed with PBS 3 times for 5 mins each and then incubated in avidin-biotin complex solution (Vectastain, Vector laboratories). They were again washed with PBS 3 times, then incubated with DAB peroxidase substrate working solution (Vector laboratories) to visualize α-Synuclein immunoreactivity. Graded alcohol was used to counter-stain, air dry, and dehydrate the sections, which were cleared in the xylene solution, followed by mounting with DPX mountant, and images were acquired using a microscope (Leica).

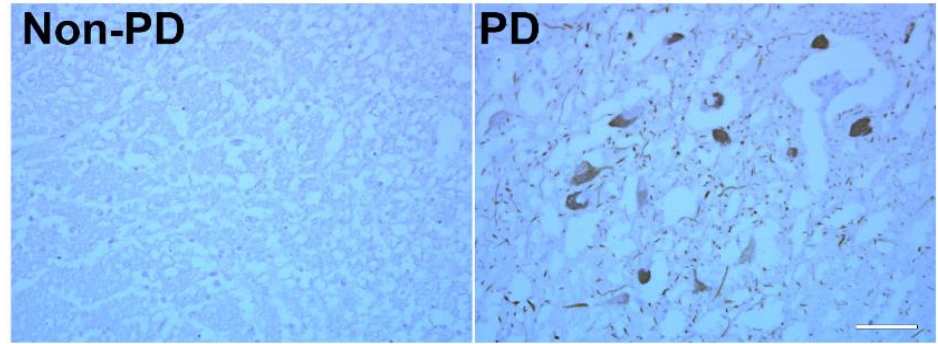

**Figure 5:** These are the representative images of human non-PD and PD brain tissues from substantia nigra from the midbrain stained for α-Synuclein. PD image shows prominent reactivity of α-Synuclein (brown staining), while the non-PD image shows the absence of it, confirming the presence of α-Synuclein immunoreactivity in PD brain.

We utilized an immunohistochemical approach to determine the α-Synuclein expression in the midbrain of PD and non-PD brains. The above images show immunoreactivity of the α-Synuclein in the neurons in the SN of the midbrains in the PD group (Fig. 5). In non-PD, those neurons did not show any positive staining.

*3.3.2. Immunofluorescence study of dopaminergic neurodegeneration*

For immunofluorescence, sections from PD and non-PD brains were subjected to antigen retrieval. After antigen retrieval, sections from the SN region were incubated with a blocking buffer, followed by incubation of a Tyrosine hydroxylase primary antibody (Millipore) overnight at 4°C. After 3 PBS washes, sections were then incubated for 2 hours with AlexaFluor-488 goat anti-rabbit secondary antibody. Vectashield® mounting medium containing DAPI (H-1200, Vector Labs) was used to wash and mount those sections.

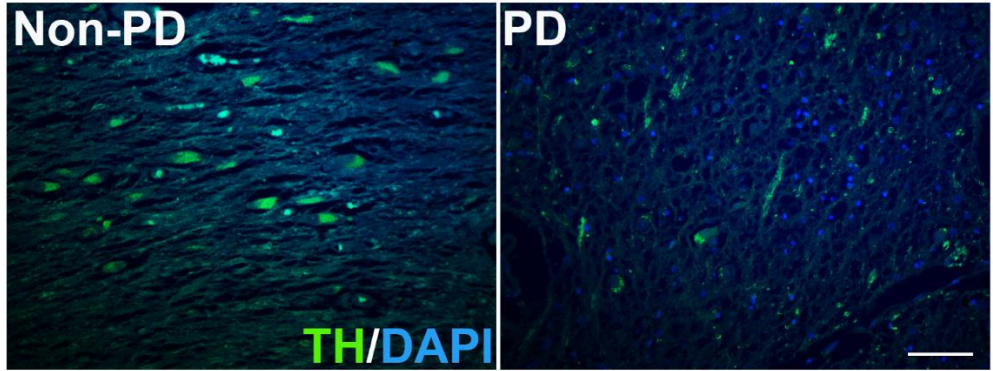

**Figure 6:** These are representative immunofluorescent-stained images of human brain tissues from the SN region. There is a prominent loss of dopaminergic neurons (green color) in PD brain tissue compared to non-PD brain tissue.

PD is characterized by the loss of dopaminergic neurons. To further examine the dopaminergic neurodegeneration in our PD and non-PD subjects, we performed immunofluorescence staining in the SN of humans using an antibody specific to Tyrosine hydroxylase (TH). Our results clearly showed a prominent loss of TH-positive neurons in the brains of PD patients compared to non-PD subjects.

## 4. Conclusion:

We applied the partial wave spectroscopy (PWS) to this work, characterized by a nanoscale sensitivity of the structural alterations and mass density changes in the tissues. In addition, to diagnose the nucleus DNA variations caused by PD, we used the IPR technique, combining it with confocal imaging. Using these dual new spectroscopy methods, it is possible to determine the intracellular structural changes of human brain tissues by identifying the disorder strength $L_d$ of the tissue's internal structures. The results show that there is a significant increase in the average and the std of $L_{d(PWS-IPR)}$ for the human brain cells and tissues of PWS and IPR analysis relative to the control. However, in the case of IPR analysis, the mean of the degree disorder value $L_d$ for the cells sample showed the most significant alterations. Notably, the disorder strength, $L_d$, measures structural changes inside the cells by detecting refractive index alterations inside the cells/tissues. Furthermore, the refractive index fluctuations are due to cell mass density fluctuation. This is explained by cytoskeletal dysregulation and Protein bundling within the affected cells, which may affect the local concentration of particles or organelles like ribosomes, DNA, and mitochondria. Finally, histological analysis of PD and non-PD tissues shows an increase in alpha-synuclein in PD brain tissues and loss of dopaminergic neurons in PD tissues than non-PD by immunohistochemistry, which are signs of PD. These depositions and neuro-degeneration change the brain structural disorder in brain tissues and change in the chromatin/DNA spatial structures, and support our results from dual optics technique.

The performance characteristics of the single biomarker $L_d$ are very promising. We provided empirical evidence that PWS and IPR can detect nanoscale architectural alterations of the PD patients' brains. It is believed that these techniques may be used as a "prescreen," identifying patients at risk who would benefit most from this approach rather than from more complex or costly tests (MRI or computerized tomography CT). It may herald the beginning of "personalized medicine" in brain disease diagnosis that employs PWS and IPR to detect wide-field brain diseases accurately.

**Author Contributions:** PP conceptualized the project; PP and MK designed the experiments; FA and IA performed PWS and IPR experiments and data analyses; JX performed histopathology experiments; FA and PP wrote the first draft; DS, IA, MK, and PP wrote the final version. All authors participated in the final version.

**Funding:** Part of this work was partially supported by the National Institutes of Health (NIH) grants (R21CA260147) to PP and R21NS128519 to MK.

**Data Availability Statement:** Data may be available to the corresponding author, PP, upon request.

**Acknowledgments:** We thank NIH for its financial support. We also acknowledge the imaging centers of Mississippi State and UTHSC for confocal imaging.

**Conflicts of Interest:** The authors declare no conflicts of interest.